\documentclass[11pt]{article}
\usepackage{graphicx,amsmath,amssymb}
\usepackage{multirow}
\begin{document}

\newtheorem{theorem}{Theorem}[section]
\newtheorem{lemma}[theorem]{Lemma}
\newtheorem{corollary}[theorem]{Corollary}
\newtheorem{proposition}[theorem]{Proposition}
\newcommand{\blackslug}{\penalty 1000\hbox{
    \vrule height 8pt width .4pt\hskip -.4pt
    \vbox{\hrule width 8pt height .4pt\vskip -.4pt
          \vskip 8pt
      \vskip -.4pt\hrule width 8pt height .4pt}
    \hskip -3.9pt
    \vrule height 8pt width .4pt}}
\newcommand{\proofend}{\quad\blackslug}
\newenvironment{proof}{$\;$\newline \noindent {Proof.}$\;$\rm}{\qad}
\newenvironment{correct}{$\;$\newline \noindent {\sc Correctness.}$\;\;\;$\rm}{\qad}
\newcommand{\qad}{\hspace*{\fill}\blackslug}
\newenvironment{definition}{$\;$\newline \noindent {\bf Definition}$\;$}{$\;$\newline}
\def\boxit#1{\vbox{\hrule\hbox{\vrule\kern4pt
  \vbox{\kern1pt#1\kern1pt}
\kern2pt\vrule}\hrule}}

\title{\vspace{-.5in}
\LARGE \bf On Scheduling Two-Stage Jobs\\
   on Multiple Two-Stage Flowshops\thanks{This work is supported in part by the National
   Natural Science Foundation of China under grants 61232001, 61472449, 61420106009, and 71221061.}}

\author{
{\sc Guangwei Wu}$^{\mbox{\footnotesize \ddag\P}}$
    \ \ \ \ \ \
{\sc Jianer Chen}$^{\mbox{\footnotesize \dag\S}}$
    \ \ \ \ \ \
{\sc Jianxin Wang}$^{\mbox{\footnotesize \textdaggerdbl}}$\footnote{Corresponding author,
        email: {\tt jxwang@csu.edu.cn}.}
\vspace*{3mm}\\
$^{\mbox{\footnotesize \ddag}}$School of Information Science and Engineering\\
    Central South University\\
    ChangSha, Hunan 410083, P.R.~China
\vspace*{2mm}\\
$^{\mbox{\footnotesize \P}}$College of Computer and Information Engineering\\
    Central South University of Forestry and Technology\\
    ChangSha, Hunan 410004, P.R.~China
\vspace*{2mm}\\
$^{\mbox{\footnotesize \dag}}$School of Computer Science \& Education Software\\ 
     Guangzhou University\\
     Guangzhou, Guangdong 510006, China
\vspace*{2mm}\\
$^{\mbox{\footnotesize \S}}$Department of Computer Science and Engineering \\
    Texas A\&M University\\
    College Station,  TX 77843, USA
}

\date{}

\maketitle

\begin{abstract}
Motivated by the current research in data centers and cloud computing, we study the problem of
scheduling a set of two-stage jobs on multiple two-stage flowshops. A new formulation for
configurations of such scheduling is proposed, which leads directly to improvements to the
complexity of scheduling algorithms for the problem. Motivated by the observation that the
costs of the two stages can be significantly different, we present deeper study on the
structures of the problem that leads to a new approach to designing scheduling algorithms for
the problem. With more thorough analysis, we show that the new approach gives very significant
improved scheduling algorithms for the problem when the costs of the two stages are different
significantly. Improved approximation algorithms for the problem are also presented.

\vspace{2mm}

\noindent {\bf keywords.}
scheduling, two-stage flowshop, pseudo-polynomial time algorithm, approximation algorithm,
cloud computing
\end{abstract}

\section{Introduction}

Scheduling is concerned with the problems of optimally allocating available resources to
process a given set of jobs. In particular, scheduling jobs on multiple machines has received
extensive study in the past four decades, in computer science, operations research, and
system sciences \cite{book16,y50}.

In this paper, we study the scheduling problems for two-stage jobs on multiple two-stage flowshops.
A machine $M$ is a {\it two-stage flowshop} (or simply a {\it flowshop}) if it consists of an
{\it $R$-processor $M_R$} and a {\it $T$-processor $M_T$} that can run in parallel. A job $J$ is
a {\it two-stage job} (or simply a {\it job}) if it consists of an {\it $R$-operation $R_J$} and
a {\it $T$-operation $T_J$} such that the $T$-operation $T_J$ cannot start on the $T$-processor
$M_T$ of a flowshop $M$ until the $R$-operation $R_J$ has been completed on the $R$-processor $M_R$
of {\it the same flowshop} $M$. When a two-stage job $J$ is {\it assigned} to a two-stage flowshop
$M$, the flowshop $M$ will first use its $R$-processor $M_R$ to process the $R$-operation $R_J$ of
$J$, then, at proper time after $M_R$ completes the processing of $R_J$, use its $T$-processor $M_T$
to process the $T$-operation $T_J$ of $J$. Thus, when we consider {\it scheduling} two-stage jobs on
multiple two-stage flowshops, we need to decide an assignment that assigns each job to a flowshop
and, for each flowshop, the execution orders of the $R$- and $T$-operations of the jobs that are
assigned to that flowshop.

Thus, the scheduling model studied in the current paper is as follows:
\begin{quote}
Given a set of $n$ two-stage jobs and a set of $m$ two-stage flowshops, construct a schedule of the
jobs on the flowshops that minimizes the makespan, i.e., the total time that elapses from the beginning
to the end for completing the execution of all the jobs.
\end{quote}

\subsection{Motivations}
\begin{figure}[h]
\centering
\includegraphics[width=0.5\linewidth]{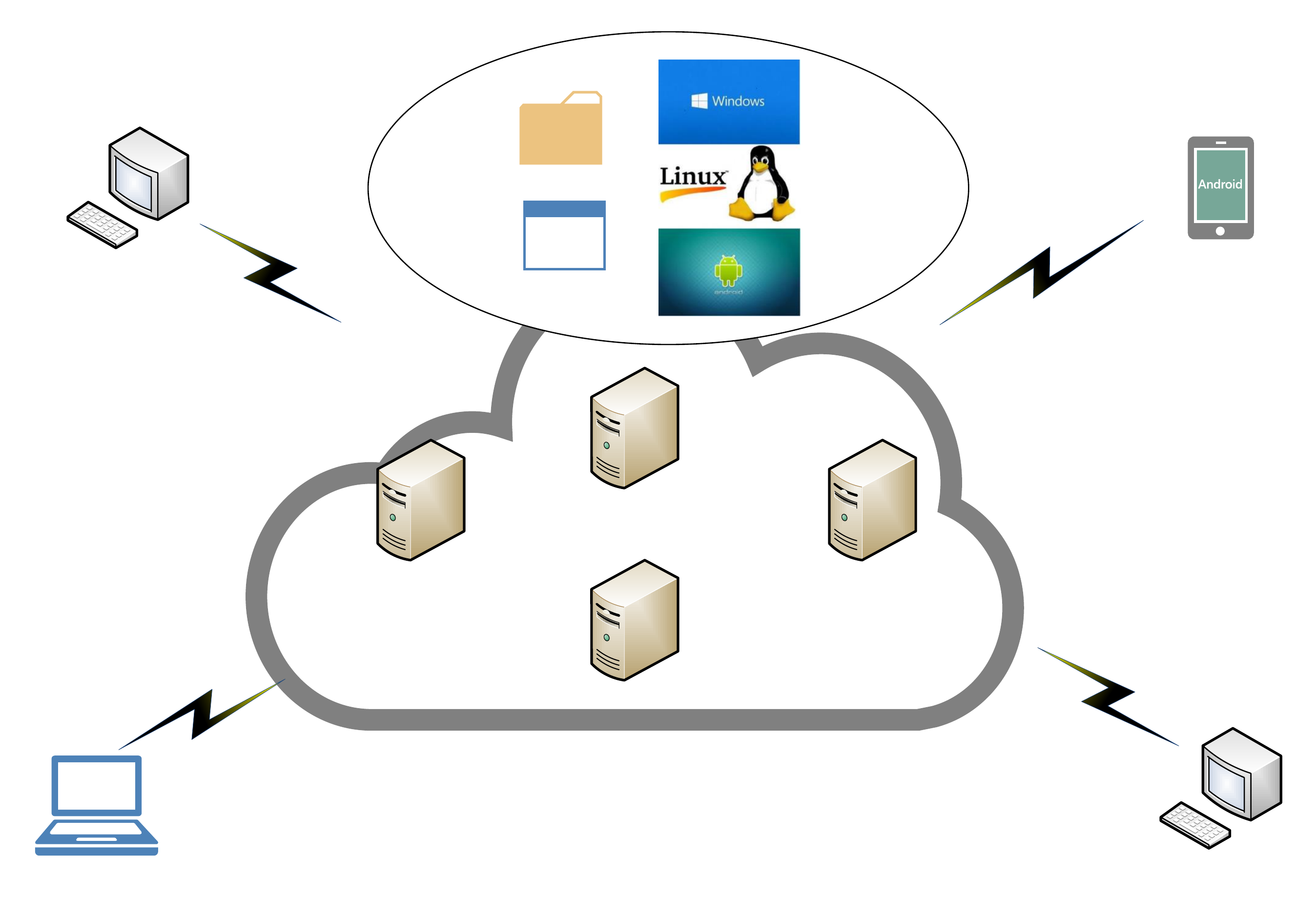}
\caption{The infrastructure of transparent computing}
\label{tcarchitecture}
\end{figure}

Our scheduling model was motivated by the current research in data centers and cloud computing. A data center is
a facility used to house servers, storage systems, and network devices, etc.~\cite{queue}. Today's data centers
contain hundreds of thousands of servers. Typical cloud computing providers rent infrastructures (IaaS), platforms
(PaaS), and softwares (SaaS) as services, while keeping the softwares and data stored in the servers in data centers.
Recently, a cloud paradigm called {\it TransCom}  \cite{transos}, based on the principle of {\it transparent computing}
\cite{transcomp}, has been proposed. This paradigm considers not only application softwares and data but also
traditional system softwares such as operation systems as resources. As a consequence, client devices in such a system
can be very light and significantly diversified, as long as they contain a small TransCom kernel and a new-generation
input/output system UEFI \cite{uefi}. Traditional operation systems, application softwares, and data are stored as
resources in the cloud. Clients dynamically request these resources selected by users, and the cloud sends  the
resources to the clients via networks. The infrastructure of such a system is shown in Figure \ref{tcarchitecture}.

In such a system, a significant amount of resources requested by clients are executable codes of system/application
softwares, which in general are large by size and commonly used by many users. Because the main memory of servers
is limited, these codes are in general stored in secondary memory such as hard disks that can be accessed by the
servers. Therefore, when a server receives a request from a client for a specific code, it will have to first read
the code from the secondary memory into the main memory, then send the code to the client via networks. As a result,
a request from a client can be divided into two operations, one is a disk-read operation $R$ that reads the requested
code/data from a secondary memory into the main memory, and the other is a network-transmission operation $T$ that sends
the code/data via the network to the requesting client. It is also natural to require that the network-transmission
operation do not start until the requested code/data has been brought into the main memory.\footnote{One may argue
that such a disk-read/network-transmission process can be done in pipeline: in this case, we can simply regard each
data block of the requested data as a ``inseparable'' job. Now for each job network-transmission must wait until
the disk-read is completed, and the job becomes two-stage. See Section 6 for more discussions.} Therefore, in such
a system, the data requests become two-stage jobs, consisting of the disk-read and the network-transmission
operations, while each server becomes a two-stage flowshop, consisting of the disk-read and network-transmission
processors (note that the disk-read and network-transmission can run in parallel in the same server), and
scheduling a given set of such requests in a multiple-server center becomes an instance of the scheduling model
we have formulated. We should remark that the time for disk-read and the time for network-transmission in a typical
server are in general comparable, and, due to the impact of cache systems, they need not to have a linear relation
\cite{virtual}. Therefore, neither can be simply ignored if we want to maintain good performance for the cloud system.

\subsection{Previous related work}

Multiple machine scheduling and flowshop scheduling have been extensively studied. We first discuss the relationship
between our scheduling model and other related scheduling models studied in the literature. Then we review the
known results specifically on our scheduling model.

First of all, the classical {\sc Makespan} problem can be regarded as scheduling one-stage jobs on multiple
one-stage machines \cite{book16}. On the other hand, scheduling two-stage jobs on a single two-stage flowshop
is the classical {\sc two-stage flowshop} problem \cite{johnson}.

Other scheduling models that deal with multiple-stage jobs include various ``hybrid'' shop scheduling problems,
such as the {\it hybrid flow shop} scheduling problem \cite{hfs99,hfs10} and the {\it hybrid/flexible job shop}
scheduling problem \cite{hjs1,hjs2}. The hybrid shop scheduling problems allow multiple machines for a stage
such that the execution of a stage operation of a job can be assigned to {\it any} machine for that stage. However,
in general there is no specific bonding requirement for the machine that executes an operation for a stage of a
job and the machine that executes the operation for the previous stage of the same job. This makes a major
difference between this model and our model: our model requires that once a job is assigned to a machine,
then the $R$-operation and the $T$-operation of the job must be executed by the $R$-processor and the
$T$-processor, respectively, of the same machine.

Indeed, in the hybrid/flexible job shop scheduling model, if each job is given a set of alternative {\it routes},
where each route is a sequence of specific flowshops, one for a stage of the job, then our scheduling problem
can be formulated as a restricted version of this very general version of the hybrid job shop scheduling problem.
However, to the authors' knowledge, this general version of the hybrid job shop scheduling problem has not
been systematically studied. Moreover, since our scheduling problem has a strong constraint that the two
stage operations of the same job be bonded to the same flowshop, a general solution to the general version of the
hybrid job shop scheduling problem will probably be not efficient and effective enough for our scheduling problem.

Another model that deals with multiple-stage jobs is that of scheduling jobs with setup costs \cite{setup},
where a job can also be regarded as a two-stage job in which one stage is the ``setup'' stage, and the other
stage is the ``regular'' processing stage. However, in the model of scheduling jobs with setup costs assumes
one-stage machines --- a machine under such a model cannot run the setup stage for one job and the regular
processing stage for another job in parallel. On the other hand, a two-stage flowshop $M$ under our model can
have its $R$-processor and $T$-processor run in parallel. Thus, when the $R$-processor of $M$ is processing the
$R$-operation for a job, the $T$-processor of $M$ can process the $T$-operation for another job at the same
time. Finally, our model is different from that of multi-processor job scheduling problem \cite{lee,chenmiranda},
where a job may require more than one processors and it holds all the requested processors during its execution.
On the other hand, a two-stage job under our current model requires the two requested processors to run one
after the other, and when one of the processors of a flowshop is running for the job, the other processor of the
flowshop may be used for processing other jobs.

Except some research directly related to specific applications, the problem of scheduling two-stage jobs on
multiple two-stage flowshops had not been studied thoroughly until very recently. He, Kusiak, and Artiba
\cite{davidhe} seem the first group who studied the problem, motivated by applications in glass manufacturing,
and proposed a heuristic algorithm. Vairaktarakis and Elhafsi \cite{elhafsi} also considered the problem in
their study on the hybrid multi-stage flowshop problem. In particular, a pseudo-polynomial time algorithm was
proposed in \cite{elhafsi} for scheduling two-stage jobs on two two-stage flowshops. Zhang and van de Velde
\cite{xzhang} presented constant ratio approximation algorithms for scheduling two-stage jobs on two and
three two-stage flowshops. Very recently,\footnote{In fact, our research is independent of \cite{dong}: we
became aware of the result of \cite{dong} only after the current paper had been completed.} following a formulation
similar to that of \cite{elhafsi}, Dong {\it et al.}~\cite{dong} proposed a pseudo-polynomial time algorithm
for scheduling two-stage jobs on $m$ two-stage flowshops for a fixed constant $m$, and developed a fully
polynomial-time approximation scheme for the problem based on the pseudo-polynomial time algorithm. We also
note that approximation algorithms for $k$-stage jobs on multiple $k$-stage flowshops for general $k$ have be
studied recently \cite{tong}.

\subsection{Our main results}

Our research in the current paper was motivated by our current project on data center and cloud computing,
as described in the previous section. Therefore, we are looking for more efficient algorithms for scheduling
two-stage jobs on multiple two-stage flowshops, which not only improve previous best theoretical complexity
bound, but also run much faster in practice.

First of all, we propose a new formulation to describe configurations for schedules of two-stage jobs on multiple
two-stage flowshops. Our formulation is very different from those studied in the literature \cite{elhafsi,dong}.
We show that dynamic programming based on our formulation directly leads to improvements on the complexity of
algorithms for scheduling two-stage jobs on multiple two-stage flowshops, in terms of both theoretical bound
and practical performance.

Our further study on the problem was motivated by the observation that in many cases in practice, the 
execution times for the two stages can differ very significantly. We present deeper study on the structures 
of the problem that leads to a more carefully designed algorithm. With more thorough analysis, we are able 
to show that the new approach will give a very significantly improved scheduling algorithm for the problem 
when the costs of the two stages are significantly different. Improved approximation algorithms for the 
problem are also presented.

The paper is organized as follows. Formal definitions and some preliminary results related to the
problem are given in Section 2. The new scheduling formulation for the problem is proposed and
improved pseudo-polynomial time exact algorithms based on the new formulation are presented in
Section 3. Section 4 is devoted to faster algorithms for the case when the execution times of the
two stages are significantly different. An improved approximation algorithm for the problem is given
in Section 5 for the problem when the number of flowshops is bounded by a constant. We conclude the
paper in Section 6 with remarks and suggested future research.

\section{Single flowshop scheduling and dual scheduling}
For $n$ two-stage jobs $J_1$, $\ldots$, $J_n$ to be processed in a system $\{M_1, \ldots, M_m\}$ of $m$ identical
two-stage flowshops, we make the following ``standard'' assumptions (variations and extensions of this model will
be discussed in Section 6
):
\begin{enumerate}
 \item each job consists of an $R$-operation and a $T$-operation;

 \vspace*{1mm}

 \item each flowshop has an $R$-processor and a $T$-processor that can run in parallel and can process the
   $R$-operations and the $T$-operations, respectively, of the jobs;

 \vspace*{1mm}

 \item the $R$-operation and $T$-operation of a job must be executed in the $R$-processor and $T$-processor,
     respectively, of the same flowshop, in such a way that the $T$-operation cannot start unless the
     $R$-operation is completed;

 \vspace*{1mm}

 \item there is no precedence constraints among the jobs; and

 \vspace*{1mm}

 \item preemption is not allowed.
\end{enumerate}

Under the model above, each job $J_i$ can be represented by a pair $(r_i, t_i)$ of integers, where $r_i$, the
{\it $R$-time}, is the time for processing the $R$-operation of $J_i$ by an $R$-processor, and $t_i$, the
{\it $T$-time}, is the time for processing the $T$-operation of $J_i$ by a $T$-processor. A {\it schedule}
$\cal S$ of a set of jobs $\{J_1, \ldots, J_n\}$ on $m$ flowshops $M_1$, $\ldots$, $M_m$ consists of an
{\it assignment} that assigns each job to a flowshop, and, for each flowshop, the execution orders of the $R$-
and $T$-operations of the jobs assigned to that flowshop in its corresponding processors. The {\it completion
time} of a flowshop $M$ under the schedule $\cal S$ is the time when $M$ finishes the execution of the last
$T$-operation for the jobs assigned to $M$ (assuming all flowshops are available at the initial time $0$).
The {\it makespan} $C_{\max}$ of $\cal S$ is the largest flowshop completion time under the schedule
$\cal S$ over all flowshops. Following the three-field notation $\alpha|\beta|\gamma$ suggested by Graham
{\it et al.}~\cite{graham}, this scheduling model can be written as $P | \mbox{2FL} | C_{\max}$, or
$P_m | \mbox{2FL} | C_{\max}$ if the number $m$ of flowshops is a fixed constant.

\subsection{Two-stage job scheduling on a single two-stage flowshop}

For $m = 1$, the problem $P_1 |\mbox{2FL}|C_{\max}$ becomes the two-stage flow shop problem. Without loss of
generality, a schedule of a set of two-stage jobs on a single two-stage flowshop can be given by an ordered
sequence $\langle J_1, J_2, \ldots, J_t \rangle$ of the jobs such that both the executions of the $R$-operations
and $T$-operations of the jobs, by the $R$-processor and $T$-processor of the flowshop, respectively, strictly
follow the given order \cite{johnson}. If our interests are in minimizing the makespan of schedules, then we
can make the following assumptions.

\begin{lemma}
\label{lem22}
Let ${\cal S} = \langle J_1, J_2, \ldots, J_t \rangle$ be a two-stage job schedule on a single two-stage
flowshop, where $J_i = (r_i, t_i)$, for $1 \leq i \leq t$. Let $\bar{\rho}_h$ and $\bar{\tau}_h$, respectively,
be the times at which the $R$-operation and the $T$-operation of job $J_h$ are started. Then for all $h$,
$1 \leq h \leq t$, we can assume:

\medskip

{\rm (1)} \  $\bar{\rho}_h = \sum_{i=1}^{h-1} r_i$; and

\smallskip

{\rm (2)} \ $\bar{\tau}_h = \max\{ \bar{\rho}_h + r_h, \bar{\tau}_{h-1} + t_{h-1} \}$.

\begin{proof}
By the assumption, both the executions of the $R$-operations and the $T$-operations of the jobs follow the
given order. Since the $R$-operation of the job $J_h$ cannot start unless the $R$-operations of all jobs $J_1$,
$\ldots$, $J_{h-1}$ are completed on the $R$-processor of the flowshop, we must have
$\bar{\rho}_h \geq \sum_{i=1}^{h-1} r_i$. If $\bar{\rho}_h > \sum_{i=1}^{h-1} r_i$, then we can let the
$R$-operation of the job $J_h$ start at time $\bar{\rho}_h' = \sum_{i=1}^{h-1} r_i$. Note that this change
does not delay any other process --- in particular, since the $T$-operation of $J_h$ starts at time
$\bar{\tau}_h$, which must be at least $\bar{\rho}_h + r_i$. Now since the $R$-operation of $J_h$ starts at
time $\bar{\rho}_h' = \sum_{i=1}^{h-1} r_i$ and finishes at time $\bar{\rho}_h' + r_i < \bar{\rho}_h + r_i$,
the $T$-operation of $J_h$ can still start at time $\bar{\tau}_h  \geq \bar{\rho}_h + r_i > \bar{\rho}_h' + r_i$.
For all other jobs, since the starting and finishing times of their $R$-operations and $T$-operations are
unchanged, the schedule remains a valid schedule, with no change in the completion time of the flowshop.
Applying this process repeatedly, we can fill all ``gaps'' in the execution of the $R$-processor of the flowshop
(i.e., the idle time in the $R$-processor of the flowshop between the finish of the $R$-operation of a job and
the start of the $R$-operation of the next job). The result is a valid schedule of the jobs, with no change
in the completion time of the flowshop, and satisfies the condition $\bar{\rho}_h = \sum_{i=1}^{h-1} r_i$ for
all $1 \leq h \leq t$. This proves (1).

The proof of (2) is simple: the $R$-operation of the job $J_h$ is finished at time $\bar{\rho}_h + r_h$, and
the $T$-operation of the job $J_{h-1}$ is finished at time $\bar{\tau}_{h-1} + t_{h-1}$. Therefore, at time
$\max\{ \bar{\rho}_h + r_h, \bar{\tau}_{h-1} + t_{h-1} \}$, the $T$-operation of the job $J_h$ can always start,
with no reason to further wait if our objective is to minimize the completion time of the flowshop. Moreover,
this is the earliest time at which the $T$-operation of $J_h$ can start.
\end{proof}
\end{lemma}

Scheduling two-stage jobs on a single two-stage flowshop, i.e., the two-stage flow shop problem
$P_1|\mbox{2FL}|C_{\max}$, can be solved optimally in time $O(n \log n)$ using the classical Johnson's
algorithm. In terms of our model, Johnson's algorithm can be described as follows  (for more details,
see \cite{johnson}):
\begin{quote}
  {\bf Johnson's Algorithm \cite{johnson}.}

  Given a set of two-stage jobs $(r_i, t_i)$, $1 \leq i \leq n$, divide the jobs into two disjoint groups
  $G_1$ and $G_2$, where $G_1$ contains all jobs $(r_h, t_h)$ with $r_h \leq t_h$, and $G_2$ contains all
  jobs $(r_g, t_g)$ with $r_g > t_g$. Order the jobs in a sequence such that the first part consists of the
  jobs in $G_1$, sorted in nondecreasing order of $R$-times, and the second part consists of the jobs in
  $G_2$, sorted in nonincreasing order of $T$-times. The schedule using the order of this sequence minimizes
  the completion time of the flowshop over all schedules of the jobs on the flowshop.
\end{quote}

{\it Johnson's order} of a set of two-stage jobs is to order the jobs into a sequence that satisfies the conditions
given by Johnson's Algorithm above. Therefore, once we determined how the jobs are assigned to the flowshops,
Johnson's order of the jobs assigned to each flowshop will give an optimal execution order for the flowshop. As a
result, what that remains unsolved is how we determine the assignment of the jobs to the flowshops. Unfortunately,
this task is intractable. In fact, in the special case where the $R$-time of every job is $0$, the problem becomes
the classical {\sc Makespan} problem $P | | C_{\max}$, where we are asked to optimally schedule a set of (one-stage)
jobs on a set of identical (one-stage) machines. The {\sc Makespan} problem is NP-hard for two machines \cite{bruno74},
and is strongly NP-hard for three or more machines \cite{gj}. As a consequence, our problem $P_m|\mbox{2FL}|C_{\max}$
is NP-hard when $m \geq 2$ and NP-hard in the strong sense when $m \geq 3$.

Johnson's orders of jobs on all flowshops can be constructed by a single sorting process on the input job set, as
given by the following lemma, whose proof is straightforward thus is omitted.

\begin{lemma}
\label{lem21}
If a job sequence $\cal S$ satisfies Johnson's order, then every subsequence of $\cal S$ also satisfies Johnson's
order.
\end{lemma}

Therefore, if we first sort the input job set in Johnson's order, which can obviously be done in time $O(n \log n)$,
then pick the jobs in that order and assign them to flowshops, then every flowshop receives a subset of jobs in
their Johnson's order, which directly gives the optimal execution order of the job subset on the flowshop. In the
rest of this paper, we will always assume that any sequence of jobs in our consideration is in Johnson's order,
unless we explicitly indicate otherwise.

Lemma~\ref{lem22} indicates that in an optimal schedule on a single flowshop based on Johnson's order, we can
simply follow Johnson's order and let the $R$-processor of the flowshop consecutively execute the $R$-operations
of the jobs without idle time until all $R$-operations are completed, and start immediately the $T$-operation of
the job $J_h$ as soon as the $R$-operation of $J_h$ and the $T$-operation of the job $J_{h-1}$ are completed. This
observation greatly helps us in dealing with two-stage jobs on multiple two-stage flowshops. In particular, for a
partial assignment of jobs on a flowshop, its corresponding (optimal) schedule now can be characterized by a
pair $(\rho, \tau)$, which gives the finish times of the $R$-operation and the $T$-operation of the last job
assigned to the flowshop. The pair $(\rho, \tau)$, which will be called the {\it status} of the schedule,
can be easily updated, based on the formulas given in Lemma~\ref{lem22}, when a new job is added to the flowshop.

\subsection{Dual jobs and dual schedules}

For a two-stage job $J_i = (r_i, t_i)$, the {\it dual job} of $J_i$ is $J_i^d = (t_i, r_i)$ (i.e., the dual
job $J_i^d$ is obtained from the original job $J_i$ by swapping its $R$- and $T$-times). Let
${\cal S} = \langle J_1, J_2, \ldots, J_n \rangle$ be a schedule of two-stage jobs on a two-stage flowshop.
The {\it dual schedule} of $\cal S$ on the dual jobs of $\cal S$ is given by
${\cal S}^d = \langle J_n^d, \ldots, J_2^d, J_1^d \rangle$, where $J_i^d$ is the dual job of $J_i$ for
$1 \leq i \leq n$. It is interesting to observe and easy to verify that if the schedule $\cal S$ follows
Johnson's order, then the dual schedule ${\cal S}^d$ also follows Johnson's order. In fact, we have a more
general result, as giving in the following theorem.

\begin{theorem}
\label{thm23}
For $1 \leq i \leq n$, let $J_i^d$ be the dual job of the two-stage job $J_i$. On a single two-stage flowshop,
the optimal schedule of the job set $G = \{ J_1, J_2, \ldots, J_n \}$ and the optimal schedule of the dual
job set $G^d = \{ J_1^d, J_2^d, \ldots, J_n^d \}$ have the same completion time. Moreover, if a schedule
${\cal S}$ is optimal for the job set $G$ then its dual schedule ${\cal S}^d$ is optimal for the dual job
set $G^d$.

\begin{proof}
For each $h$, let $J_h = (r_h, t_h)$. Thus, the dual job of $J_h$ is $J_h^d = (t_h, r_h)$. Let
${\cal S} = \langle J_1, J_2, \ldots, J_n \rangle$ be an optimal schedule for the job set $G$, where, by
Lemma~\ref{lem22}, for each job $J_h$, the $R$-operation starts at time $\bar{\rho}_h = \sum_{i=1}^{h-1} r_i$
and finishes at time $\bar{\rho}_h + r_h$, and the $T$-operation starts at time
$\bar{\tau}_h = \max\{\bar{\rho}_h + r_h, \bar{\tau}_{h-1} + t_{h-1}\}$ and finishes at time $\bar{\tau}_h + t_h$.
The completion time of the schedule $\cal S$ is $\tau^* = \bar{\tau}_n + t_n$.

Now consider the schedule ${\cal S}_1^d = \langle J_n^d, \ldots, J_2^d, J_1^d \rangle$ for the dual job set
$G^d$, where for each dual job $J_h^d = (t_h, r_h)$, $1 \leq h \leq n$, the $R$-operation of $J_h^d$ starts at
time $\bar{\rho}_h' = \tau^* - (\bar{\tau}_h + t_h)$ and finishes at time $\bar{\rho}_h'' = \tau^* - \bar{\tau}_h$,
and the $T$-operation of $J_h^d$ starts at time $\bar{\tau}_h' = \tau^* - \sum_{i=1}^h r_i$ and finishes at time
$\bar{\tau}_h'' = \tau^* - \sum_{i=1}^{h-1} r_i$ (see Figure~\ref{fig21} for an illustration).

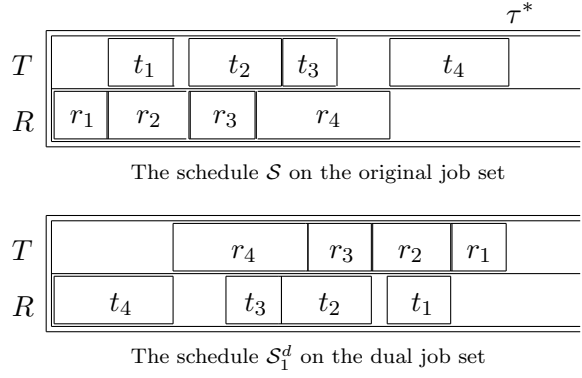
\begin{figure}
\begin{center}
\begin{picture}(205,130)
\put(21,81){\line(1,0){20}} \put(21,81){\line(0,1){18}}
\put(21,99){\line(1,0){20}} \put(41,81){\line(0,1){18}}
\put(27,87){$r_1$}
\put(41,81){\line(1,0){30}} \put(41.5,81){\line(0,1){18}}
\put(41,99){\line(1,0){30}} \put(72,81){\line(0,1){18}}
\put(52,87){$r_2$}
\put(72,81){\line(1,0){25}} \put(72.5,81){\line(0,1){18}}
\put(72,99){\line(1,0){25}} \put(97,81){\line(0,1){18}}
\put(81,87){$r_3$}
\put(98,81){\line(1,0){50}} \put(97.5,81){\line(0,1){18}}
\put(98,99){\line(1,0){50}} \put(148,81){\line(0,1){18}}
\put(120,87){$r_4$}
\put(41.5,101){\line(1,0){25}} \put(41.5,101){\line(0,1){18}}
\put(41.5,119){\line(1,0){25}} \put(66,101){\line(0,1){18}}
\put(51,106){$t_1$}
\put(72,101){\line(1,0){35}} \put(72,101){\line(0,1){18}}
\put(72,119){\line(1,0){35}} \put(107,101){\line(0,1){18}}
\put(87,106){$t_2$}
\put(107.5,101){\line(1,0){20}} \put(107.5,101){\line(0,1){18}}
\put(107.5,119){\line(1,0){20}} \put(128,101){\line(0,1){18}}
\put(113,106){$t_3$}
\put(148,101){\line(1,0){45}} \put(148,101){\line(0,1){18}}
\put(148,119){\line(1,0){45}} \put(193,101){\line(0,1){18}}
\put(168,106){$t_4$}
\put(21,11){\line(1,0){45}} \put(21,11){\line(0,1){18}}
\put(21,29){\line(1,0){45}} \put(66,11){\line(0,1){18}}
\put(42,16){$t_4$}
\put(86,11){\line(1,0){20}} \put(86,11){\line(0,1){18}}
\put(86,29){\line(1,0){20}} \put(107,11){\line(0,1){18}}
\put(93,16){$t_3$}
\put(106,11){\line(1,0){35}} \put(107,11){\line(0,1){18}}
\put(106,29){\line(1,0){35}} \put(141,11){\line(0,1){18}}
\put(121,16){$t_2$}
\put(147,11){\line(1,0){24}} \put(147,11){\line(0,1){18}}
\put(147,29){\line(1,0){24}} \put(171,11){\line(0,1){18}}
\put(156,16){$t_1$}
\put(66,31){\line(1,0){50}} \put(66,31){\line(0,1){18}}
\put(66,49){\line(1,0){50}} \put(117,31){\line(0,1){18}}
\put(88,36){$r_4$}
\put(116,31){\line(1,0){25}} \put(117,31){\line(0,1){18}}
\put(116,49){\line(1,0){25}} \put(141,31){\line(0,1){18}}
\put(125,36){$r_3$}
\put(141,31){\line(1,0){30}} \put(141.5,31){\line(0,1){18}}
\put(141,49){\line(1,0){30}} \put(171,31){\line(0,1){18}}
\put(151,36){$r_2$}
\put(171,31){\line(1,0){21}} \put(171.5,31){\line(0,1){18}}
\put(171,49){\line(1,0){21}} \put(192,31){\line(0,1){18}}
\put(177,36){$r_1$}
\put(18,8){\line(1,0){202}} \put(18,52){\line(1,0){202}}
\put(20,10){\line(1,0){200}} \put(20,50){\line(1,0){200}}
\put(20,30){\line(1,0){200}} \put(20,10){\line(0,1){40}}
\put(18,8){\line(0,1){44}}
\put(5,35){$T$} \put(5,14){$R$}
\put(50,-4){\scriptsize The schedule ${\cal S}_1^d$ on the dual job set}
\put(18,78){\line(1,0){202}} \put(18,122){\line(1,0){202}}
\put(20,80){\line(1,0){200}} \put(20,120){\line(1,0){200}}
\put(20,100){\line(1,0){200}} \put(20,80){\line(0,1){40}}
\put(18,78){\line(0,1){44}}
\put(5,105){$T$} \put(5,84){$R$} \put(192,125){$\tau^*$}
\put(50,66){\scriptsize The schedule $\cal S$ on the original job set}
\end{picture}
\end{center}

\vspace*{-4mm}

\caption{Schedules for a job set and its dual}
\label{fig21}
\end{figure}
Note that since $\bar{\tau}_{h+1} \geq \bar{\tau}_h + t_h$, the job $J_h^d$ has its $R$-operation starting
at time $\bar{\rho}_h' = \tau^* - (\bar{\tau}_h + t_h)$, which is not earlier than the finish time
$\bar{\rho}_{h+1}'' = \tau^* - \bar{\tau}_{h+1}$ of the $R$-operation of the job $J_{h+1}^d$. Similarly, the
job $J_h^d$ has its $T$-operation starting at time $\bar{\tau}_h' = \tau^* - \sum_{i=1}^h r_i$, which is not
earlier than (actually, is equal to) the finish time $\bar{\tau}_{h+1}'' = \tau^* - \sum_{i=1}^h r_i$ of the
$T$-operation of the job $J_{h+1}^d$. Finally, since $\bar{\tau}_h \geq \bar{\rho}_h + r_h = \sum_{i=1}^h r_i$,
the starting time $\bar{\tau}_h' = \tau^* - \sum_{i=1}^h r_i$ of the $T$-operation of the job $J_h^d$ is not
earlier than the finish time $\bar{\rho}_h'' = \tau^* - \bar{\tau}_h$ of the $R$-operation of the same job
$J_h^d$. This shows that ${\cal S}_1^d$ is a valid schedule for the dual job set $G^d$. Since the last job
$J_1^d$ in the schedule ${\cal S}_1^d$ finishes at time $\bar{\tau}_1'' = \tau^*$, the completion time of
${\cal S}_1^d$ is $\tau^*$. Now, by Lemma~\ref{lem22}, we can convert the schedule ${\cal S}_1^d$ into the
standard dual schedule ${\cal S}^d$ of $\cal S$, without increasing the completion time, where ${\cal S}^d$
is a schedule for the dual job set $G^d$ and satisfies the conditions in Lemma~\ref{lem22}.

By our assumption, $\cal S$ is an optimal schedule for the job set $G$ and has completion time $\tau^*$. Thus,
the fact that the completion time of the dual schedule ${\cal S}^d$ for the dual job set $G^d$ is not larger
than $\tau^*$ implies that the completion time of an optimal schedule for the job set $G$ is not smaller than
that of an optimal schedule for the dual job set $G^d$.

For the other direction, we start with an optimal schedule ${\cal S}^d$ for the dual job set $G^d$. Using exactly
the same procedure, we can show that the schedule $({\cal S}^d)^d$ that is dual to ${\cal S}^d$ for the job set
$(G^d)^d$ that is dual to $G^d$ has its completion time not larger than that of ${\cal S}^d$. Since the job set
$(G^d)^d$ that is dual to the dual job set $G^d$ is just the original job set $G$, this shows that the completion
time of an optimal schedule for the dual job set $G^d$ is not smaller than that of an optimal schedule for the
original job set $G$.

Combining these results, we conclude that the optimal schedule of the job set $G$ and the optimal schedule of
the dual job set $G^d$ have the same completion time. This implies that if the completion time of an optimal
schedule $\cal S$ for the job set $G$ is $\tau^*$, then the completion time of the dual schedule ${\cal S}^d$
for the dual job set $G^d$ is also $\tau^*$. Thus, ${\cal S}^d$ must be optimal for the dual job set $G^d$.
\end{proof}
\end{theorem}

Now consider scheduling two-stage jobs on multiple two-stage flowshops. Let $G = \{J_1, J_2, \ldots, J_n \}$ be
a set of two-stage jobs, and let $G^d = \{J_1^d, J_2^d, \ldots, J_n^d \}$ be the dual job set, where for each
$h$, $J_h^d$ is the dual job of the job $J_h$.

\begin{theorem}
\label{thm24}
On multiple two-stage flowshops, the optimal schedule of the job set $G$ and the optimal schedule of the dual
job set $G^d$ have the same makespan. Moreover, an optimal schedule for the job set $G$ can be easily obtained
from an optimal schedule for the dual job set $G^d$.

\begin{proof}
Suppose that $\cal S$ is an optimal schedule of the job set $G$ on $m$ two-stage flowshops, where for each
$i$, $1 \leq i \leq m$, $\cal S$ assigns a subset $G_i$ of jobs in $G$ to the $i$-th flowshop. Without loss of
generality, we can assume that $\cal S$ optimally schedules the jobs in $G_i$ on the $i$-th flowshop. Now for
each $i$, replace the schedule of $G_i$ on the $i$-th flowshop by its dual schedule for the dual job set
$G_i^d$. This gives a schedule ${\cal S}^d$ on the $m$ flowshops for the dual job set $G^d$. By Theorem~\ref{thm23},
under the schedule ${\cal S}^d$, the completion time for each flowshop is the same as that under the schedule
$\cal S$. Thus, the makespan of the schedule ${\cal S}^d$ for the dual job set $G^d$ on the $m$ flowshops is the
same as that of the schedule $\cal S$ for the job set $G$. This proves that the makespan of an optimal schedule
for $G$ on the $m$ flowshops is not smaller than that of an optimal schedule for $G^d$.

Conversely, starting with an optimal schedule of the dual job set $G^d$ on $m$ two-stage flowshops, we can
similarly construct a schedule for the original job set $G$ whose makespan is equal to that of the optimal
schedule for $G^d$, which implies that the makespan of an optimal schedule for $G^d$ on the $m$ flowshops
is not smaller than that of an optimal schedule for $G$.

Combining these results shows that the optimal schedule of the job set $G$ and the optimal schedule of the
dual job set $G^d$ have the same makespan. The discussion also explains that starting with an optimal schedule
${\cal S}^d$ for the dual job set $G^d$, by replacing the schedule on each flowshop with its dual schedule
we can obtain an optimal schedule of the job set $G$.
\end{proof}
\end{theorem}

Theorem~\ref{thm24} provides flexibility when we work on scheduling two-stage jobs on multiple two-stage
flowshops: sometimes working on the job set that is dual to the given input job set may have certain advantages.
In this case, we can simply work on the dual job set, whose optimal solutions can be easily converted into
optimal solutions for the original job set. This property will be used in Section 4.

\section{Pseudo-polynomial time algorithms for $P_m | \mbox{2FL} | C_{\max}$}

In this section, we study the problem $P_m | \mbox{2FL} | C_{\max}$, i.e., the problem of scheduling
two-stage jobs on $m$ two-stage identical flowshops, where $m$ is a fixed constant. Our input
is a set of two-stage jobs $G = \{J_1, J_2, \ldots, J_n\}$, where for each $i$, $J_i = (r_i, t_i)$, and
we are looking for a schedule of the jobs on $m$ identical two-stage flowshops $M_1$, $\ldots$, $M_m$,
that minimizes the makespan. Let $R_0 = \sum_{i=1}^n r_i$, $T_0 = \sum_{i=1}^n t_i$, and for
$0 \leq k \leq n$, let $G_k = \{J_1, J_2, \ldots, J_k\}$ be the set of the first $k$ jobs in $G$.

With a preprocessing, we can assume that the sequence $\langle J_1, J_2, \ldots, J_n \rangle$ is in Johnson's
order. If we pick the jobs in this order and assign them to the flowshops, then, by Lemma~\ref{lem21}, the
subsequence received by each flowshop $M_h$ is also in Johnson's order, which thus gives an optimal schedule
of the jobs assigned to the flowshop $M_h$. Therefore, the status of the flowshop $M_h$ at any moment can be
represented by a pair $(\rho_h, \tau_h)$ for the corresponding schedule, where $\rho_h$ and $\tau_h$
are the completion times of the $R$-processor and the $T$-processor, respectively, of the flowshop $M_h$. By
Lemma~\ref{lem22}, the status $(\rho_h, \tau_h)$ of the flowshop $M_h$ can be easily updated when a new job
$(r, t)$ is added to the flowshop $M_h$: the new completion time of the $R$-processor will be $\rho_h + r$,
and the new completion time of the $T$-processor will be $\max\{\rho_h + r, \tau_h\} + t$. For each schedule
$\cal S$ of the job subset $G_k$, the tuple $(k; \rho_1, \tau_1, \ldots, \rho_m, \tau_m)$ will be called the
{\it configuration} of $\cal S$ if under the schedule $\cal S$ for $G_k$, the status of the flowshop $M_h$
is $(\rho_h, \tau_h)$, for all $h$.

The key observation, which can be easily verified, is that for each $k > 0$, we have:

\begin{quote}

{\bf Fact A.} \ The tuple $(k; \rho_1, \tau_1, \ldots, \rho_m, \tau_m)$ is a configuration of a schedule
for the job subset $G_k$ if and only if there is a flowshop $M_d$ such that the tuple
$(k-1; \rho_1', \tau_1', \ldots, \rho_m', \tau_m')$ is a configuration of a schedule for the job subset
$G_{k-1}$, where for $i \neq d$, $\rho_i' = \rho_i$, $\tau_i' = \tau_i$, and $\rho_d'$ and $\tau_d'$ satisfy
$\rho_d = \rho_d' + r_k$ and $\tau_d = \max\{\rho_d' + r_k, \tau_d'\} + t_k$, i.e., the schedule given by
$(k; \rho_1, \tau_1, \ldots, \rho_m, \tau_m)$ is obtained by adding the job $J_k$ to flowshop $M_d$ in the
schedule given by $(k-1; \rho_1', \tau_1', \ldots, \rho_m', \tau_m')$.

\end{quote}

Fact A suggests a dynamic programming algorithm that starts with the tuple $(0; 0, 0, \ldots, 0, 0)$,
which corresponds to the unique schedule for the initial empty job subset $G_0$, and applies Fact A
repeatedly to construct all possible configurations for the schedules for the given job set $G = G_n$.
Moreover, the value $\max_h \{\tau_h\}$ for a configuration $(n; \rho_1, \tau_1, \ldots, \rho_m, \tau_m)$
gives the makespan of the schedule described by the configuration. Therefore, The configuration
$(n; \rho_1, \tau_1, \ldots, \rho_m, \tau_m)$ with $\max_h \{\tau_h\}$ being minimized over all configurations
gives a schedule for the job set $G$ on the $m$ flowshops whose makespan is the minimum over all schedules
of the job set $G$.

It is easy to see that for all $1 \leq h \leq m$, the value $\rho_h$ is an integer bounded between $0$ and
$R_0$, and the value $\tau_h$ is an integer bounded between $0$ and $R_0 + T_0$. Therefore, a straightforward
implementation of the dynamic programming algorithm runs in time $O(n m^2R_0^m (R_0+T_0)^m)$, which will be quite
significant when the values of $R_0$ and $T_0$ are large. In the following, we study how the complexity of the
algorithm is improved.

Let $(\rho_h, \tau_h)$ be the status of the flowshop $M_h$. By definition, we always have $\rho_h \leq \tau_h$.
Moreover, for any job $J_i$ assigned to the flowshop $M_h$, by Lemma~\ref{lem22}, if the $T$-operation of
$J_i$ starts no earlier than $\rho_h$, then it can always start immediately after the $T$-operation of the
previous job assigned to $M_h$ is completed, i.e., there is no ``gap'' in the execution of the $T$-processor
of $M_h$ after time $\rho_h$. This gives $\tau_h - \rho_h \leq T_0$. This observation suggests that we can
use the pair $(\rho_h, \delta_h)$ instead of the pair $(\rho_h, \tau_h)$, where $\delta_h = \tau_h - \rho_h$,
and $0 \leq \delta_h \leq T_0$. Note that the pair $(\rho_h, \tau_h)$ can be easily obtained from the pair
$(\rho_h, \delta_h)$.

Therefore, for a configuration $(k; \rho_1, \tau_1, \ldots, \rho_m, \tau_m)$ for a schedule for the job
subset $G_k$, we will represent it by the tuple $(k; \rho_1, \delta_1, \ldots, \rho_m, \delta_m)$, where
for all $h$, $\delta_h = \tau_h - \rho_h$ with $0 \leq \delta_h \leq T_0$, which will be called the
{\it s-configuration} of the schedule.

\medskip

\noindent {\bf Remark.} Our configurations and s-configurations defined above are very different from those
proposed in the literature \cite{elhafsi,dong}, where a configuration is defined based on the makespan of
the schedule (see \cite{elhafsi,dong} for more details). We will show that based on the formulation of
our configurations, much faster algorithms can be developed for the $P_m | \mbox{2FL} | C_{\max}$ problem.

\medskip

Our next improvement is based on reducing the dimension of the s-configurations. Let
${\cal S}_k = (k; \rho_1, \delta_1, \ldots, \rho_m, \delta_m)$ be an s-configuration for the job subset
$G_k$. Let $R_0^k = \sum_{i=1}^k r_i$. By Lemma~\ref{lem22}, there is no ``gap'' in the execution of the
$R$-processors of the flowshops. Therefore, $\sum_{h=1}^m \rho_h = R_0^k$. This gives

\begin{quote}

{\bf Fact B.} The value $\rho_1$ can be computed from the values $\rho_2$, $\ldots$, $\rho_m$:
$\rho_1 = R_0^k - \sum_{h=2}^m \rho_h$.

\end{quote}

Let ${\cal S}_k = (k; \rho_1, \delta_1, \rho_2, \delta_2, \ldots, \rho_m, \delta_m)$ and
${\cal S}_k' = (k; \rho_1, \delta_1', \rho_2, \delta_2, \ldots, \rho_m, \delta_m)$ be s-configurations
for the job subset $G_k$ that only differ in the completion time of the $T$-processor of flowshop $M_1$, with
$\delta_1 < \delta_1'$. It is easy to see that if we can assign the rest of the jobs $J_{k+1}$, $\ldots$,
$J_n$ to ${\cal S}_k'$ to build a minimum makespan schedule for the entire job set $G$, then the same way of
assigning the jobs $J_{k+1}$, $\ldots$, $J_n$ to ${\cal S}_k$ will also give a minimum makespan schedule of
the job set $G$. Therefore, when all other parameters are identical, we really only have to record the smallest
completion time (thus the smallest value $\delta_1$) for the $T$-processor of the flowshop $M_1$.

This suggests that we can represent all ``useful'' s-configurations for $G_k$ by a $(2m-1)$-dimensional array $H$
such that
\[ H[k; \rho_2, \delta_2, \ldots, \rho_m, \delta_m] = (\delta_1, d), \]
if by letting $\rho_1 = R_0^k - \sum_{h=2}^m \rho_h$, the value $\delta_1$ is the smallest $\delta_1'$ such
that $(k; \rho_1, \delta_1', \rho_2, \delta_2, \ldots, \rho_m, \delta_m)$ is a valid s-configuration for the job
subset $G_k$.

Now we are ready for our algorithm, which is given in Figure~\ref{new2}.
\begin{figure}[htbp]
\setbox4=\vbox{\hsize28pc \noindent\strut
\vspace{1mm} \footnotesize {\bf Algorithm DynProg-I}\\
{\sc input}:  a set $G = \{J_1, \ldots, J_n \}$ of two-stage jobs, in Johnson's order\\
{\sc output}: an optimal schedule of $G$ on $m$ two-stage flowshops

1. \hspace*{1mm} {\bf for} all $0 \leq k \leq n$, $0 \leq \rho_h \leq R_0$,
      $0 \leq \delta_h \leq T_0$, $2 \leq h \leq m$ {\bf do}\\
\hspace*{11mm}      $H[k; \rho_2, \delta_2, \ldots, \rho_m, \delta_m] = (+\infty, 0)$;\\
2. \hspace*{1mm} $H[0; 0, 0, \ldots, 0, 0] = (0, 0)$; \\
3. \hspace*{1mm} {\bf for} $k = 0$ {\bf to} $n-1$ {\bf do} \\
3.1 \hspace*{2mm} {\bf for} each $H[k, \rho_2, \delta_2, \ldots, \rho_m, \delta_m] = (\delta_1, d_k)$
       with $\delta_1 \neq +\infty$  {\bf do}\\
3.2 \hspace*{6mm} $\rho_1 = R_0^k - \sum_{h=2}^m \rho_h$;\\
3.3 \hspace*{6mm} {\bf for} $d = 1$ {\bf to} $m$ {\bf do} \\
3.4  \hspace*{10mm} {\bf for} $(1 \leq h \leq m)\; \&\;(h \neq d)$ {\bf do}
    \{ $\rho_h' = \rho_h$; $\delta_h' = \delta_h$; \} \\
3.5 \hspace*{10mm} $\rho_d' = \rho_d + r_{k+1}$; \ \
               $\delta_d' = \max\{r_{k+1}, \delta_d\} + t_{k+1} - r_{k+1}$;\\
3.6 \hspace*{10mm} {\bf if} $H[k+1; \rho_2', \delta_2', \ldots, \rho_m', \delta_m'] = (\delta_1, d_{k+1})$
                 with $\delta_1' < \delta_1$ \\
3.7 \hspace*{10mm} {\bf then} $H[k+1; \rho_2', \delta_2', \ldots, \rho_m', \delta_m'] = (\delta_1', d)$; \\
4. \hspace*{1mm} return the $H[n; \rho_2, \delta_2, \ldots, \rho_m, \delta_m] = (\delta_1, d_n)$ that minimized
      the value \\
   \hspace*{5mm}  $\max_{1 \leq h \leq m} \{\rho_h + \delta_h\}$.
  \vspace{1mm} \strut} $$\boxit{\box4}$$
 \vspace{-12mm}
\caption{An improved algorithm for $P_m | \mbox{2FL} | C_{\max}$} \label{new2}
\end{figure}

We give some explanations for the algorithm. Steps 3.4-3.7 add the job $J_{k+1}$ to the
$d$-th flowshop in the schedule for the job subset $G_k$ with an s-configuration
${\cal S} = (k; \rho_1, \delta_1, \rho_2, \delta_2, \ldots, \rho_m, \delta_m)$. Thus, before adding
the job $J_{k+1}$, the completion times of the $R$-processor and the $T$-processor of the $d$-th
flowshop are $\rho_d$ and $\rho_d + \delta_d$, respectively. By Lemma~\ref{lem22}, after adding the job
$J_{k+1}$, the completion time of the $R$-processor becomes $\rho_d' = \rho_d + r_{k+1}$, and the
completion time of the $T$-processor is $\max\{ \rho_d + r_{k+1}, \rho_d + \delta_d\} + t_{k+1}$. Therefore,
by the definition, after adding the job $J_{k+1}$, we should have
\begin{eqnarray*}
 \delta_d' & = & (\max\{ \rho_d + r_{k+1}, \rho_d + \delta_d\} + t_{k+1}) - \rho_d' \\
    & = & \max\{r_{k+1}, \delta_d\} + \rho_d + t_{k+1} - (\rho_d + r_{k+1})\\
    & = & \max\{r_{k+1}, \delta_d\} + t_{k+1} - r_{k+1},
\end{eqnarray*}
as shown in step 3.5 of the algorithm.

Note that the last row $H[n; \**, \ldots, \**]$ of the $(2m-1)$-dimensinal array $H$ includes all
possible s-configurations of the schedules for the job set $G = G_k$ on the $m$ flowshops. Moreover,
the value $\max_{1 \leq h \leq m} \{\rho_h + \delta_h\}$ for an element
$H[n; \rho_2, \delta_2, \ldots, \rho_m, \delta_m] = (\delta_1, d)$ (where $\rho_1 = R_0 - \sum_{h=2}^m \rho_h$)
gives the makespan of the schedule described by $H[n; \rho_2, \delta_2, \ldots, \rho_m, \delta_m] = (\delta_1, d)$.
Therefore, the one with $\max_{1 \leq h \leq m} \{\rho_h + \delta_h\}$ being minimized over all
$H[n; \rho_2, \delta_2, \ldots, \rho_m, \delta_m] = (\delta_1, d)$ with $\delta_1 \neq +\infty$, as the
one returned in step 4 of the algorithm, gives a schedule for the job set $G$ on the $m$ flowshops whose
makespan is the minimum over all schedules of the job set $G$. According to the algorithm, the value
$H[k; \rho_2, \delta_2, \ldots, \rho_m, \delta_m] = (\delta_1, d)$ also records that the last job $J_k$ in
the job subset $G_k$ was added to the flowshop $M_d$ to obtain the s-configuration corresponding to
$H[k; \rho_2, \delta_2, \ldots, \rho_m, \delta_m] = (\delta_1, d)$. With this information, the actual schedule
corresponding to the element $H[k; \rho_2, \delta_2, \ldots, \rho_m, \delta_m] = (\delta_1, d)$ can be
re-constructed as follows: (1) if $d \neq 1$, then look through $H[k-1; \rho_2, \delta_2, \ldots, \rho_{d-1},
\delta_{d-1}, \rho_d - r_k, \delta_d', \rho_{d+1}, \delta_{d+1}, \ldots, \rho_m, \delta_m] = (\delta_1, d)$
with $\delta_1 \neq +\infty$ for all $0 \leq \delta_d' \leq T_0$; and (2) if $d = 1$, then look at 
the element $H[k-1; \rho_2, \delta_2, \ldots, \rho_m, \delta_m] = (\delta_1', d)$, we will find an element 
$H[k-1; \rho_2', \delta_2', \ldots, \rho_m', \delta_m']$ for the job subset $G_{k-1}$ that, when $J_k$ is 
added to the flowshop $M_d$, gives the array element $H[k; \rho_2, \delta_2, \ldots, \rho_m, \delta_m]$.
Now with this array element $H[k-1; \rho_2', \delta_2', \ldots, \rho_m', \delta_m']$ for $G_{k-1}$, we 
will find where the job $J_{k-1}$ went and what is the corresponding array element for $G_{k-2}$, and so 
on. Thus, starting from the array element returned in step 4 of the algorithm {\bf DynProg-I}, we will be 
able to re-construct an optimal schedule for the job set $G$.

Since we have $0 \leq k \leq n$, and $0 \leq \rho_h \leq R_0$, $0 \leq \delta_h \leq T_0$, for all
$2 \leq h \leq m$, the $(2m-1)$-dimensional array $H$ has a size $O(nR_0^{m-1}T_0^{m-1})$. The algorithm
basically goes through the array $H$, element by element, and applies steps 3.2-3.7 on each element,
which take time $O(m^2)$. Thus, the algorithm takes time $O(n m^2 R_0^{m-1} T_0^{m-1})$ and space
$O(n R_0^{m-1} T_0^{m-1})$ (i.e., the space for the array $H$). Note that if we want to re-construct the
optimal schedule based on the element returned in step 4 of the algorithm, we can go through the rows 
$H[k; \**, \ldots, \**]$ of the array $H$ (i.e., the first index of the array) backwards (i.e., $k$ goes 
from $n$ to $1$), as we described above. This will
take additional $O(nm T_0)$ time. Now we are ready to conclude the algorithm with the following theorem.

\begin{theorem}
\label{thm31}
An optimal schedule for $n$ two-stage jobs on $m$ two-stage flowshops can be constructed in time
$O(n m^2 R_0^{m-1} T_0^{m-1})$ and space $O(n R_0^{m-1} T_0^{m-1})$.
\end{theorem}

We compare Theorem~\ref{thm31} with the existing result given in \cite{dong}, which is the only known result
for the $P_m | \mbox{2FL} | C_{\max}$ problem. The algorithm given in \cite{dong} is based on a very different
definition for configurations for schedules of two-stage jobs on $m$ two-stage flowshops, and has running
time $O(nm^2(R_0 + T_0)^{2m-1})$ and space $O(m(R_0 + T_0)^{2m-2})$. Therefore, in terms of the running time,
our algorithm in Theorem~\ref{thm31} not only replaces the larger factor $R_0 + T_0$ by smaller factors $R_0$
and $T_0$, but also reduces the exponent from $2m-1$ to $2m-2$. In terms of the space complexity, our algorithm
seems to use more space because in general $n > m$. However, a careful examination shows that the algorithm
given in \cite{dong} seems to only return the value of the makespan of an optimal schedule without giving
the actual optimal schedule. In order to also return an actual schedule, the algorithm in \cite{dong}
seems to have to increase its space complexity to at least $O(n m (R_0 + T_0)^{2m-2})$. On the other
hand, if we are only interested in the value of the makespan of an optimal schedule for the given job
set, then we can modify our algorithm to run in space $O(R_0^{m-1} T_0^{m-1})$: according to the algorithm
{\bf DynProg-I}, each row $H[k+1; \**, \ldots, \**]$ of the array $H$ is computed based solely on the
previous row $H[k; \**, \ldots, \**]$. Therefore, we only need to keep two rows of the array $H$, and
repeatedly compute the next row based on the current row. This will use space $O(R_0^{m-1} T_0^{m-1})$,
which also improves the space complexity of the algorithm in \cite{dong}. In conclusion, our algorithm
in Theorem~\ref{thm31} improves both time complexity and space complexity of the algorithm given in
\cite{dong}.

\section{Dealing with the case when $R_0$ and $T_0$ differ significantly}

In certain cases in practice, the values $R_0$ and $T_0$ can differ very significantly. Consider
the situation in data centers as we described in Section 1. In order to improve the process of
data-read/network-transformation, severs in the center may keep certain commonly used software codes
in the main memory so that the time-consuming process of data-read can be avoided (see, for example,
\cite{virtual}). Thus, client requests for the code will become two-stage jobs $J_i = (r_i, t_i)$
with $r_i = 0$. As a consequence, the value $R_0 = \sum_{i=1}^n r_i$ can be significantly smaller
than the value $T_0 = \sum_{i=1}^n t_i$. On the other hand, certain data centers may consist of a
large number of slow-speed servers (e.g., PC's) but equipped with high-speed networks \cite{transos},
which may make $T_0$ much smaller than $R_0$.

In this section, we will study how to reduce the sizes of the dimensions of the configurations for
the schedules in the case where the values $R_0$ and $T_0$ differ very significantly. This will lead
to significant improvements on the complexity of scheduling algorithms. We divide the study into two
cases: (1) $T_0$ is significantly larger than $R_0$ (i.e., $T_0 \gg R_0$), and (2) $R_0$ is significantly
larger than $T_0$ (i.e., $T_0 \ll R_0$). We first consider the case $T_0 \gg R_0$.

Since all flowshops are identical, we can arbitrarily re-order the flowshops. In particular, we can
order the flowshops so that the completion times of the $R$-operations of the flowshops are non-increasing.
We call an s-configuration $(k; \rho_1, \delta_1, \ldots, \rho_m, \delta_m)$ {\it canonical} if
$\rho_1 \geq \rho_2 \geq \cdots \geq \rho_m$. Any s-configuration of a schedule can be converted into
a canonical s-configuration by properly re-ordering the flowshops. Therefore, we only need to consider
canonical s-configurations.

Let $(k; \rho_1, \delta_1, \ldots, \rho_m, \delta_m)$ be a canonical s-configuration for a schedule
${\cal S}_k$ for the job subset $G_k$. By Lemma~\ref{lem22}, there is no ``gap'' in the execution of the
$R$-processors of the flowshops, so $\sum_{h=1}^m \rho_i \leq R_0$. This gives reduced upper bounds for
the completion time of the $R$-processors of the flowshops:

\begin{quote}

{\bf Fact C.} In a canonical s-configuration $(k; \rho_1, \delta_1, \ldots, \rho_m, \delta_m)$,
$\rho_h \leq R_0/h$, for all $1 \leq h \leq m$.

\end{quote}

As we explained in the previous section, if our objective is to minimize the makespan, then when all the
values $k$, $\rho_1$, $\rho_2$, $\delta_2$, $\ldots$, $\rho_m$, $\delta_m$ are given, we only need to record
the smallest $\delta_1'$ such that $(k; \rho_1, \delta_1', \rho_2, \delta_2, \ldots, \rho_m, \delta_m)$
corresponds to a valid schedule for the job subset $G_k$. This reduces the number of dimensions for
the s-configurations by $1$.\footnote{However, unlike algorithm {\bf DynProg-I}, we will not be able to
remove the dimension for $\rho_1$ in s-configurations. This will become clearer in our discussion.}

In contrast to Fact C, the values $\delta_h$ can be very large (recall $T_0 \gg R_0$). We now consider
how to deal with the situations when the values $\delta_h$ are large.

Fix an $h$, and consider the $h$-th flowshop. By Fact C, the completion time of the $R$-processor of
the flowshop can never be larger than $R_0/h \leq R_0$. If the completion time $\rho_h + \delta_h $ of
the $h$-th flowshop is larger than or equal to $R_0$, then for any further job $J_p$ assigned to the
flowshop, the $T$-operation of $J_p$ can always start immediately when the $T$-processor is available.
Therefore, all further jobs assigned to the flowshop can have their $T$-operations executed consecutively
with no execution ``gaps'' in the $T$-processor of the flowshop.\footnote{Actually by Fact C, this
statement holds true for the $h$-th flowshop when $\rho_h + \delta_h \geq R_0/h$. However, since later
we may need to re-order the flowshops to keep the s-configurations canonical, the $h$-th flowshop may
become the $h'$-the flowshop with $\rho_h + \delta_h < R_0/h'$. Thus, here we pick the looser but
more universal bound $\rho_h + \delta_h \geq R_0$ that is independent of $h$ and also simplifies our
discussion.} Thus, the completion time of the flowshop will only depend on the $T$-operations of the
further assigned jobs, while is independent of the $R$-operations of these jobs. We can use a single
value $\rho_h = R_0/h + 1$ to record this situation so that the pair $(R_0/h+1, \delta_h)$ represents
a real status $(\rho_h', \delta_h')$ of the flowshop where $\rho_h' + \delta_h' \geq R_0$, and
$\delta_h = \rho_h' + \delta_h' - R_0$. Note that when a new job $J_p = (r_p, t_p)$ is added to the
flowshop, the corresponding pair of the flowshop is simply changed to $(R_0/h+1, \delta_h + t_p)$.

This observation enables us to represent the status of the $h$-th flowshop by a pair $(\rho_h, \delta_h)$,
where either $0 \leq \rho_h \leq R_0/h$ and $0 \leq \rho_h + \delta_h < R_0$ (which implies
$0 \leq \delta_h < R_0$), or $\rho_h = R_0/h+1$ and $0 \leq \delta_h \leq T_0$ (which implies that the
completion time for the $T$-processor of the flowshop is $R_0 + \delta_h$). A pair is a {\it valid pair}
for the $h$-th flowshop if it satisfies these conditions. The total number of valid pairs for the $h$-th
flowshop is bounded by $(R_0/h + 1)R_0 + (T_0 + 1) = O(R_0^2/h + T_0)$. Note that all valid pairs can be
given by a two-dimensional array (i.e., a matrix) with $R_0/h + 2$ rows in which each of the first
$R_0/h + 1$ rows contains $R_0$ elements and the last row contains $T_0 + 1$ elements (if you like, you
can also regard this matrix as an $(R_0/h+1) \times R_0$ matrix plus a one-dimensional array of size $T_0 + 1$).

Summarizing the above discussions, we conclude that all ``useful'' canonical s-configurations for the job
subset $G_k$, for all $k$, can be represented by a $(2m)$-dimensional array $H'$ whose elements are
$(m+1)$-tuples, such that if
\[ H'[k; \rho_1, \rho_2, \delta_2, \ldots, \rho_m, \delta_m] = (d_k, \delta_1', \rho_2', \ldots, \rho_m'), \]
where $0 \leq \rho_1 \leq R_0$, and for $2 \leq h \leq m$, $(\rho_h, \delta_h)$ is a valid pair for the $h$-th flowshop,
then there is a canonical s-configuration $(k; \rho_1', \delta_1', \rho_2', \delta_2', \ldots, \rho_m', \delta_m')$
for a valid schedule for the job subset $G_k$, where $\rho_1' = \rho_1$ and $\delta_1'$ is the smallest when all
other parameters satisfy their conditions, such that for each $h$, $2 \leq h \leq m$,

\medskip

(1) if $\rho_h \leq R_0/h$, then $\rho_h + \delta_h < R_0$, $\rho_h' = \rho_h$ and $\delta_h' = \delta_h$, and

(2) if $\rho_h = R_0/h + 1$, then $\rho_h' + \delta_h' \geq R_0$, and $\delta_h = \rho_h' + \delta_h' - R_0$.

\medskip

Finally the value $d_k$ in the array element, $1 \leq d_k \leq m$, indicates that the last job $J_k$ in the
job subset $G_k$ is assigned to the $d_k$-th flowshop.

Note that in the case $\rho_h = R_0/h + 1$, there can be many different values for $\rho_h'$ that thus
correspond to many different canonical s-configurations that satisfy the above conditions. As
explained earlier, in this case, different choices of the values $\rho_h'$ will not affect the
makespan of the final schedule of the job set $G$. Thus, we can pick any valid values (not necessarily
the smallest) for these $\rho_h'$, as long as their sum plus $\rho_1$ is equal to $\sum_{i=1}^k r_i$.

Since the total number of valid pairs for the $h$-th flowshop, for $2 \leq h \leq m$, is $O(R_0^2/h + T_0)$,
and $0 \leq k \leq n$, we conclude that the number of elements in the array $H'$ is bounded by
\[ O((n+1) (R_0+1) \prod_{h=2}^m (R_0^2/h + T_0)) = O(n (R_0^{2m-1}/m! + R_0 T_0^{m-1})). \]
Finally, since each element of $H'$ is an $(m+1)$-tuple, we conclude that the array $H'$ takes space
$O(nm(R_0^{2m-1}/m! + R_0 T_0^{m-1}))$.

We explain how to extend a schedule for the job subset $G_k$ to a schedule for the job subset
$G_{k+1}$ when the job $J_{k+1}$ is added. For this, suppose that we have a canonical s-configuration
${\cal S}_k = (k; \rho_1', \delta_1', \rho_2', \delta_2', \ldots, \rho_m', \delta_m')$
for $G_k$ that is given by the element of the array $H'$:
\[ H'[k; \rho_1, \rho_2, \delta_2, \ldots, \rho_m, \delta_m] = (d_k, \delta_1', \rho_2', \ldots, \rho_m'), \]
as explained above. Note that the s-configuration ${\cal S}_k$ can be completely re-constructed when the
corresponding element of $H'$ is given:

\medskip

(1) \hspace*{1mm} the status $(\rho_1', \delta_1')$ for the first flowshop is $(\rho_1, \delta_1')$;

(2) \hspace*{1mm} for $2 \leq h \leq m$,

(2.1)\ if $0 \leq \rho_h \leq R_0/h$, then the status $(\rho_h', \delta_h')$ of the $h$-th flowshop is
    $(\rho_h, \delta_h)$; and

(2.2)\ if $\rho_h = R_0/h+1$, then the status $(\rho_h', \delta_h')$ of the $h$-th flowshop is
    $(\rho_h', R_0 + \delta_h - \rho_h')$.

\medskip

\noindent Note that in case (2.2), what matter is that the completion time of the $T$-processor is
equal to $\rho_h' + (R_0 + \delta_h - \rho_h') = R_0 + \delta_h$, while the value $\rho_h'$ may vary
as long as it satisfies $\rho_1 + \sum_{h=2}^m \rho_h' = R_0$.

Now suppose that we decide to add the job $J_{k+1} = (r_{k+1}, t_{k+1})$ to the $d$-th flowshop in the
s-configuration ${\cal S}_k$. Then the resulting configuration for the job subset $G_{k+1}$ will become
\[  (k+1; \rho_1'', \delta_1'', \rho_2'', \delta_2'', \ldots, \rho_m'', \delta_m''), \]
where $\rho_d'' = \rho_d' + r_{k+1}$ and $\delta_d'' = \max\{r_{k+1}, \delta_d'\} + t_{k+1} - r_{k+1}$
(see the explanation given for algorithm {\bf DynProg-I} in the previous section), and for $h \neq d$,
$\rho_h'' = \rho_h'$ and $\tau_h'' = \tau_h'$. This, after properly sorting the flowshops using the
values of $\rho_h''$, becomes a canonical s-configuration
\[ {\cal S}_{k+1} = (k; \bar{\rho}_1', \bar{\delta}_1', \bar{\rho}_2', \bar{\delta}_2',
       \ldots, \bar{\rho}_m', \bar{\delta}_m') \]
for the job subset $G_{k+1}$. Assume the $d$-th flowshop in ${\cal S}_k$ becomes the $d_{k+1}$-th flowshop
in ${\cal S}_{k+1}$. Now let $\bar{\rho}_1 = \bar{\rho}_1'$, and for each $h$, $2 \leq h \leq m$, if
$\bar{\rho}_h' + \bar{\delta}_h' < R_0$ then let $\bar{\rho}_h = \bar{\rho}_h'$ and
$\bar{\delta}_h = \bar{\delta}_h'$, and if $\bar{\rho}_h' + \bar{\delta}_h' \geq R_0$ then let
$\bar{\rho}_h = R_0/h + 1$ and $\bar{\delta}_h = \bar{\rho}_h' + \bar{\delta}_h' - R_0$. With these values,
look at the array element
\[ H'[k+1; \bar{\rho}_1, \bar{\rho}_2, \bar{\delta}_2, \ldots, \bar{\rho}_m, \bar{\delta}_m].  \]
If the element has not been assigned a value, yet, then assign it the value
$(d_{k+1}, \bar{\delta}_1', \bar{\rho}_2', \ldots, \bar{\rho}_m')$. If the element already has a value
$(d', \bar{\delta}_1'', \bar{\rho}_2'', \ldots, \bar{\rho}_m'')$ but $\bar{\delta}_1' < \bar{\delta}_1''$,
then change its value to $(d_{k+1}, \bar{\delta}_1', \bar{\rho}_2', \ldots, \bar{\rho}_m')$. This completes
the process of extending the canonical s-configuration given by the array element
$H'[k, \rho_1, \rho_2, \delta_2, \ldots, \rho_m, \delta_m]$, when job $J_{k+1}$ is added to the $d$-th
flowshop, to an array element for a canonical s-configuration for the job subset $G_{k+1}$. It is easy
to see that this process takes time $O(m)$.

Using the above description to replace the steps 3.1-3.7 in the algorithm {\bf DynProg-I} gives the procedure of
extending a canonical s-configuration for $G_k$ to a canonical s-configuration for $G_{k+1}$. This, plus certain
obvious modifications in other steps, gives a new algorithm {\bf DynProg-II} for the $P_m|\mbox{2FL}|C_{\max}$
problem. Since the number of elements of the array $H'$ is bounded by $O(n (R_0^{2m-1}/m! + R_0 T_0^{m-1}))$,
we conclude that the time complexity of the algorithm {\bf DynProg-II} is $O(n m^2(R_0^{2m-1}/m! + R_0 T_0^{m-1}))$.
Similarly as we explained for the algorithm {\bf DynProg-I}, once we apply algorithm {\bf DynProg-II} and
find the array element of $H'$ that gives a minimum makespan schedule of the job set $G$ on the $m$ flowshops,
we can use the array to construct the actual schedule by backtracking the array, row by row, in the same
amount of time.

Now we describe how to deal with job sets $G$ when $T_0 \ll R_0$. Let $G^d$ be the dual job set of $G$, and
let $R_0'$ and $T_0'$ be the sums of the times of the $R$-operations and of the $T$-operations, respectively, of
the jobs in $G^d$. By the definition, $R_0' = T_0$ and $T_0' = R_0$. Therefore, we have $T_0' \gg R_0'$. Thus,
applying the algorithm {\bf DynProg-II} on the dual job set $G^d$ will construct an optimal schedule for $G^d$
in time $O(nm^2 ((R_0')^{2m-1}/m! + R_0' (T_0')^{m-1})) = O(nm^2 (T_0^{2m-1}/m! + T_0 R_0^{m-1}))$. By Theorem~\ref{thm24},
an optimal schedule for the job set $G$ can be easily constructed from the optimal schedule for the dual job
set $G^d$ returned by the algorithm {\bf DynProg-II}.

This allows us to close this section with the following theorem:

\begin{theorem}
\label{thm32}
An optimal schedule for a set of two-stage jobs $\{J_1, \ldots, J_n\}$ on $m$ two-stage flowshops, where
$J_k = (r_k, t_k)$, can be constructed in time $O(nm^2 (T_{\min}^{2m-1}/m! + T_{\min} T_{\max}^{m-1}))$ and
space $O(nm (T_{\min}^{2m-1}/m! + T_{\min} T_{\max}^{m-1}))$, where $T_{\min}$ and $T_{\max}$ are the smaller
and the larger, respectively, of the values $\sum_{k=1}^n r_k$ and $\sum_{k=1}^n t_k$.
\end{theorem}

When $T_{\max} \gg T_{\min}$, Theorem~\ref{thm32} provides significant improvements. For example, if
$T_{\max} = T_{\min}^2$, then, for a fixed constant $m$, the time complexity of the algorithm given
in Theorem~\ref{thm32} is of the order $O(n T_{\min}^{2m-1}) = O(n T_{\max}^{m-1/2})$, which almost
matches the best pseudo-polynomial time algorithm for the {\sc Makespan} problem $P_m | | C_{\max}$
on $m$ machines \cite{book16}, which can be regarded as a much simpler version of the
$P_m | \mbox{2FL}| C_{\max}$ problem in which all jobs are one-stage jobs and all machines are
one-stage flowshop. On the other hand, the time complexity of algorithm {\bf DynProg-I} given in the
previous section is of the order $O(n T_{\min}^{3m-3}) = O(n T_{\min}^{m-1} T_{\max}^{m-1})$.

\section{Approximation algorithms for $P_m | \mbox{2FL} | C_{\max}$}

Based on the well-known techniques in approximation algorithms \cite{ik75}, we can use the
pseudo-polynomial time algorithms given in previous sections to develop approximation algorithms
for the problem $P_m | \mbox{2FL} | C_{\max}$. We present such approximation algorithms in this
section.

Since the problem $P_1 | \mbox{2FL} | C_{\max}$ of scheduling two-stage jobs on a single
two-stage flowshop can be solved optimally in polynomial time, we will assume $m \geq 2$
in our discussion in this section.

Let $\cal S$ be a schedule for a set $G = \{J_1, J_2, \ldots, J_n\}$ of two-stage jobs on $m$
two-stage flowshops. The schedule $\cal S$ can be described by a partition of the job set $G$ into
$m$ subsets, which can be given by a {\it job index partition} $(P_1, P_2, \ldots, P_m)$, which is
a disjoint partition of $\{1,2, \ldots, n\}$. Thus, under the schedule $\cal S$, the job subset
$G_h = \{J_k \mid k \in P_h\}$ of $G$ is assigned to the $h$-th flowshop $M_h$, for all $h$. The
schedule $\cal S$ can be easily implemented if for each $h$, we order the jobs in $G_h$ in Johnson's
order. For each $h$, let $C({\cal S}, h)$ be the completion time of the flowshop $M_h$ under the
schedule $\cal S$. Thus, the {\it makespan} $C_{\max}({\cal S})$ of the schedule $\cal S$ is equal
to $\max_{1 \leq h \leq m} \{C({\cal S}, h)\}$. We say that the schedule $\cal S$ {\it achieves
its makespan on the flowshop $M_h$} if $C_{\max}({\cal S}) = C({\cal S}, h)$.

The following lemma will be useful in the analysis of our approximation algorithms.

\begin{lemma}
\label{lem51}
Let ${\cal S}_k = \langle J_1, J_2, \ldots, J_k \rangle$ be a schedule on a two-stage flowshop
$M$. If we replace each job $J_i = (r_i, t_i)$ with the job $J_i' = (r_i+1, t_i+1)$ in the
schedule ${\cal S}_k$, then the completion time $\tau_k'$ of the resulting schedule
${\cal S}_k' = \langle J_1', J_2', \ldots, J_k' \rangle$ is bounded by $k+1$ plus the completion
time $\tau_k$ of ${\cal S}_k$.

\begin{proof}
We prove the lemma by induction on $k$. For $k=1$, the lemma holds true since it is easy to
see that increasing both the $R$-time and $T$-time of the job $J_1$ increases the completion
time of the single-job schedule for $\{J_1\}$ by at most $2$.

Now consider the case $k > 1$. Consider the ``partial'' schedule
${\cal S}_{k-1} = \langle J_1, J_2, \ldots, J_{k-1} \rangle$ which is obtained by taking off
the last job $J_k$ from the schedule ${\cal S}_k$. Let the completion times of the $R$-processor
and the $T$-processor of the flowshop $M$ under the schedule ${\cal S}_{k-1}$ be $\rho_{k-1}$ and
$\tau_{k-1}$, respectively. By Lemma~\ref{lem22}, $\tau_k=\max\{ \rho_{k-1}+r_k, \tau_{k-1}\}+t_k$.

Now replace each job $J_i = (r_i, t_i)$ in the schedule ${\cal S}_{k-1}$ with the job $J_i' = (r_i+1, t_i+1)$,
for $1 \leq i \leq k-1$. By the inductive hypothesis, the completion time $\tau_{k-1}'$ of the
resulting schedule ${\cal S}_{k-1}' = \langle J_1', J_2', \ldots, J_{k-1}' \rangle$ is bounded by
$\tau_{k-1} + (k-1) + 1 = \tau_{k-1} + k$. Again by Lemma~\ref{lem22}, the completion time $\rho_{k-1}'$
of the $R$-processor of $M$ on the schedule ${\cal S}_{k-1}'$ is equal to $\rho_{k-1} + (k-1)$. Now we
can add the job $J_k' = (r_k+1, t_k+1)$ to the schedule ${\cal S}_{k-1}'$ to obtain the schedule
${\cal S}_k' = \langle J_1', J_2', \ldots, J_k' \rangle$. By Lemma~\ref{lem22}, the completion time
$\tau_k'$ of the flowshop $M$ under the schedule ${\cal S}_k'$ is
\begin{eqnarray*}
  \tau_k' & = & \max\{\rho_{k-1}' + (r_k + 1), \tau_{k-1}' \} + (t_k + 1) \\
     & \leq & \max\{\rho_{k-1} + (k-1) + (r_k + 1), \tau_{k-1} + k \} + (t_k + 1) \\
     & = & \max\{\rho_{k-1} + r_k + k, \tau_{k-1} + k \} + (t_k + 1)\\
     & = & (\max\{\rho_{k-1} + r_k, \tau_{k-1} \} + t_k) +  (k + 1)\\
     & = & \tau_k + (k + 1).
\end{eqnarray*}
This completes the proof of the lemma.
\end{proof}
\end{lemma}

Now we are ready to present the approximation algorithm, which is given in Figure~\ref{new4}.
\begin{figure}[htbp]
\setbox4=\vbox{\hsize28pc \noindent\strut
\vspace{1mm} \footnotesize {\bf Algorithm Approx}\\
{\sc input}:  a set $G = \{J_1, \ldots, J_n \}$ of two-stage jobs, where $J_k = (r_k, t_k)$
     for all $k$, and $\epsilon > 0$\\
{\sc output}: a schedule of $G$ on $m$ identical two-stage flowshops

1. let $T_{\max} = \max\{R_0, T_0\}$ and $K = \epsilon \cdot T_{\max}/(n m)$; \\
2. {\bf for} $i = 1$ {\bf to} $n$ {\bf do} \
       \{ $r_i' = \lfloor r_i/K \rfloor$; \ $t_i' = \lfloor t_i/K \rfloor$ \};\\
3. let $G' = \{J_1', \ldots, J_n'\}$, where for each $i$, $J_i' = (r_i', t_i')$;\\
4. apply an algorithm $\cal A$ on $G'$, assuming $\cal A$ returns an optimal schedule ${\cal S}'$ \\
\hspace*{3mm} for $G'$, given by a job index partition $(P_1, \ldots, P_m)$;\\
5. return the schedule $\cal S$ for $G$ that uses the same job index partition\\ \hspace*{3mm} $(P_1, \ldots, P_m)$.
 \vspace{1mm} \strut} $$\boxit{\box4}$$
 \vspace{-9mm}
\caption{An approximation algorithm for $P_m | \mbox{2FL} | C_{\max}$} \label{new4}
\end{figure}

We first study how well the schedule $\cal S$ returned by the algorithm can approximation the
optimal schedule for the job set $G$ on the $m$ flowshops.

Both the schedule ${\cal S}'$ for the job set $G'$ and the schedule $\cal S$ for the job set $G$
use the same job index partition $(P_1, \ldots, P_m)$. Suppose that $\cal S$ achieves its makespan
$C_{\max}({\cal S})$ on flowshop $M_h$ and that ${\cal S}'$ achieves its makespan $C_{\max}({\cal S}')$
on flowshop $M_{h'}$. Let ${\cal S}_0$ be an optimal schedule for the job set $G$ that has a job
index partition $(P_1', \ldots, P_m')$ and achieves its makespan $C_{\max}({\cal S}_0)$ on flowshop
$M_d$. Let ${\cal S}_0'$ be the schedule for the job set $G'$ that also uses the job index partition
$(P_1', \ldots, P_m')$ and achieves its makespan $C_{\max}({\cal S}_0')$ on flowshop $M_{d'}$.

We need some further notations for our analysis. As we defined, for a schedule $\cal S$ based on the
job index partition $(P_1, \ldots, P_m)$ and a flowshop $M_h$, $C({\cal S}, h)$ denotes the completion
time of the flowshop $M_h$ under the schedule $\cal S$. Let $K$ be the number defined in step 1 of
the algorithm. We will use the notation $C({\cal S}/K, h)$ to denote the completion time of the flowshop
$M_h$ under the schedule $\cal S$ with each $J_k = (r_k, t_k)$ of the jobs in the job subset
$\{J_i \mid i \in P_h\}$ replaced by the job $(r_k/K, t_k/K)$, i.e., we shrink each of the jobs by a
factor $K$. Note here the jobs may no longer have integral $R$-time and $T$-time -- this will not affect
the complexity of our algorithms and the correctness of our analysis because we will only use this
notation and our algorithms will not take advantage of this relaxation. With these notations, we have
\begin{eqnarray}
C_{\max}({\cal S}) & = & C({\cal S}, h) = K \cdot C({\cal S}/K, h) \label{eqq1} \\
  & \leq & K \cdot C({\cal S}', h) + Kn \leq K \cdot C({\cal S}', h') + Kn \label{eqq2}\\
  & \leq & K \cdot C({\cal S}_0', d') + Kn \leq K \cdot C({\cal S}_0/K, d') + Kn \label{eqq3}\\
  & = & C({\cal S}_0, d') + Kn \leq C({\cal S}_0, d) + Kn \label{eqq4}\\
  & = & Opt(G) + Kn  \label{eqq5}
\end{eqnarray}

We explain the derivations in (\ref{eqq1})-(\ref{eqq5}). The first equality in (\ref{eqq1}) is
because by our assumption, the schedule $\cal S$ achieves its makespan on flowshop $M_h$. The
second equality in (\ref{eqq1}) is obvious: if we proportionally shrink the $R$-time and the
$T$-time of each job in flowshop $M_h$ by a factor $K$, then the completion time of the flowshop
$M_h$ is also shrunk by a factor $K$.

Now consider (\ref{eqq2}). To simplify the notations without loss of generality, let 
$\langle J_1, J_2, \ldots, J_k \rangle$ be the schedule on the flowshop $M_h$ induced from the 
schedule $\cal S$. Since there are $m \geq 2$ flowshops, we have $k \leq n-1$. The schedule 
${\cal S}/K$ on the ``shrunk'' jobs induces a schedule $\langle J_1/K, J_2/K, \ldots, J_k/K \rangle$ 
on the flowshop $M_h$, where $J_p/K = (r_p/K, t_p/K)$, $1 \leq p \leq k$. If we replace each 
shrunk job $J_p/K = (r_p/K, t_p/K)$ in the flowshop $M_h$ by the job 
$(J_p/K)^+ = (\lfloor r_p/K \rfloor + 1, \lfloor t_p/K \rfloor + 1)$ with larger $R$-time and
$T$-time, the completion time $C(({\cal S}/K)^+, h)$ of the flowshop $M_h$ under the resulting
schedule $({\cal S}/K)^+ = \langle (J_1/K)^+, \ldots, (J_k/K)^+ \rangle$ will not be decreased. 
That is, $C({\cal S}/K, h) \leq C(({\cal S}/K)^+, h)$. On the other hand, the schedule
$({\cal S}/K)^+$ on the flowshop $M_h$ can be obtained from the schedule ${\cal S}/K$, via the
schedule ${\cal S}'$ defined in step 4 of the algorithm {\bf Approx}, as follows: first we 
replace in ${\cal S}/K$ each shrunk job $J_p/K = (r_p/K, t_p/K)$ in $M_h$ by the job 
$J_p' = (\lfloor r_p/K \rfloor, \lfloor t_p/K \rfloor)$, as defined in step 3 of the algorithm
{\bf Approx}. Since neither of the $R$-time and $T$-time of each job is increased, the resulting
schedule ${\cal S}'$, as given in step 4 of the algorithm that shares the same job index partition
with $\cal S$, has its completion time on the flowshop $M_h$ not larger than that of ${\cal S}/K$.
That is, $C({\cal S}', h) \leq C(({\cal S}/K), h)$. Now the schedule $({\cal S}/K)^+$ on the
flowshop $M_h$ is obtained from the schedule ${\cal S}'$ by increasing both $R$-time and $T$-time
of each job in $M_h$ by $1$. By Lemma~\ref{lem51},
$C(({\cal S}/K)^+, h) \leq C({\cal S}', h) + (k+1) \leq C({\cal S}', h) + n$ (here we have used the
fact $k \leq n-1$). This, combined with $C({\cal S}/K, h) \leq C(({\cal S}/K)^+, h)$ proved above,
gives immediately $C({\cal S}/K, h) \leq C({\cal S}', h) + n$. This proves the first inequality
in (\ref{eqq2}). The second inequality in (\ref{eqq2}) is because we assume the schedule ${\cal S}'$
achieves its makespan on flowshop $M_{h'}$.

By the algorithm {\bf Approx}, ${\cal S}'$ is an optimal schedule for the job set $G'$ that achieves 
its makespan on flowshop $M_{h'}$. By our assumption, ${\cal S}_0'$ is also a (not necessarily optimal) 
schedule for the job set $G'$ that achieves its makespan on flowshop $M_{d'}$. This explains the first 
inequality in (\ref{eqq3}). The second inequality in (\ref{eqq3}) is based on the observation that if 
we replace the $R$-time and $T$-time of each job $J_p' = (\lfloor r_p/K \rfloor, \lfloor t_p/K \rfloor)$ 
in flowshop $M_{d'}$ by not smaller numbers $r_p/K$ and $t_p/K$, respectively, the completion time of 
the flowshop $M_{d'}$ would not decrease.

The reason for the equality in (\ref{eqq4}) is the same as that for the second equality in (\ref{eqq1}).
The inequality in (\ref{eqq4}) is because the schedule ${\cal S}_0$ achieves its makespan on flowshop
$M_d$. Finally, the equality in (\ref{eqq5}) is because we assume ${\cal S}_0$ is an optimal schedule
for the job set $G$ (here we have used $Opt(G)$ for the makespan of an optimal schedule for the job
set $G$).

According to the derivation in (\ref{eqq1})-(\ref{eqq5}), $C_{\max}({\cal S}) \leq Opt(G) + Kn$. From
$K = \epsilon\cdot T_{\max}/(nm)$, we get $Kn = \epsilon \cdot T_{\max}/m \leq \epsilon \cdot Opt(G)$,
where the inequality is based on the obvious fact $T_{\max}/m \leq Opt(G)$. This gives us the following
relation:
\begin{equation}
   C_{\max}({\cal S}) \leq Opt(G) (1 + \epsilon) \mbox{\hspace*{5mm} or \hspace*{5mm}}
    C_{\max}({\cal S})/Opt(G) \leq 1 + \epsilon.    \label{eqq6}
\end{equation}

The time complexity of the algorithm {\bf Approx} depends on the algorithm $\cal A$ we use
in step 4 of the algorithm to construct the optimal schedule ${\cal S}'$ for the job set
$G' = \{J_1', J_2', \ldots, J_n'\}$. For example, if we use the algorithm {\bf DynProg-I} in
Figure~\ref{new2}, then by Theorem~\ref{thm31}, the running time of the algorithm {\bf DynProg-I},
thus the running time of the algorithm {\bf Approx} will be $O(n m^2 (R_0')^{m-1}(T_0')^{m-1})$,
where
\[ R_0' = \sum_{i=1}^n r_i'
     = \sum_{i=1}^n \lfloor r_i/K \rfloor \leq \frac{1}{K} \left( \sum_{i=1}^n r_i \right)
     = \left( \sum_{i=1}^n r_i \right) \cdot \frac{nm}{\epsilon \cdot T_{\max}}
     \leq \frac{nm}{\epsilon}, \]
here we have used the inequality $\sum_{i=1}^n r_i = R_0 \leq T_{\max}$. Similarly,
$T_0' \leq (nm)/\epsilon$. This shows that the running time of the algorithm {\bf Approx} is
bounded by $O(n^{2m-1}m^{2m}/\epsilon^{2m-2})$. This concludes the discussion of this
section with the following theorem.

\begin{theorem}
\label{ftas}
There is an algorithm for the $P_m | \mbox{2FL} | C_{\max}$ problem
that on a set $G$ of $n$ two-stage jobs and any real number $\epsilon > 0$, constructs a schedule
for the job set $G$ on $m$ two-stage flowshops with a makespan bounded by $Opt(G)(1 + \epsilon)$.
Moreover, the running time of the algorithm is $O(n^{2m-1}m^{2m}/\epsilon^{2m-2})$.
\end{theorem}

Compared to the approximation algorithm given in \cite{dong}, which also produces a schedule with
makespan bounded by $Opt(G)(1 + \epsilon)$ but runs in time $O(2^{2m-1}n^{2m}m^{2m+1}/\epsilon^{2m-1})$,
our algorithm in Theorem~\ref{ftas} gives an obvious improvement on the running time.

When $m$ is a fixed constant, the algorithm in Theorem~\ref{ftas} runs in time polynomial in $n$
and $1/\epsilon$. In the literature of approximation algorithms, such approximation algorithms with
a ratio $1 + \epsilon$ are called {\it fully polynomial-time approximation schemes} (FPTAS) \cite{gj}.
Thus, Theorem~\ref{ftas} claims that the $P_m | \mbox{2FL} | C_{\max}$ problem has an FPTAS when
$m$ is a fixed constant.

\section{Conclusion}

Motivated by the current research in data centers and cloud computing, we studied the scheduling
problem of two-stage jobs on multiple two-stage flowshops, which in particular addresses the
scheduling issues of data transmissions between clients and servers in data centers in the cloud
computing framework based on the principle of transparent computing. The problem is NP-hard.
Pseudo-polynomial time algorithms for the problem were presented that produce optimal solutions
for the problem when the number of flowshops is a fixed constant. In particular, with thorough
analysis, we show that for certain cases, much faster pseudo-polynomial time algorithms can be
achieved. Approximation algorithms for the problem have also been developed and studied. Our
algorithms improve previous known algorithms for the problem.

Needs and considerations in cloud computing practice suggest many further research topics that require
the study of variations and extensions of our scheduling model. We list some of them below for future
research.

A cloud computing center may have many servers with different powers, ranging from large mainframe
computers to small PC's. The disks connected to the servers and the network bandwidth available for
the servers can also differ. Moreover, the disk-read on a server at some moment may even not be needed
if the requested data is already in the server's main memory. This calls for the study of scheduling
two-stage jobs on heterogeneous two-stage flowshops. Our scheduling model can be easily extended to
include this situation: suppose that we need to schedule $n$ jobs $J_1$, $\ldots$, $J_n$ on $m$ flowshops
$M_1$, $\ldots$, $M_m$ that may not be identical, then we can represent each two-stage job $J_i$ by
$m$ pairs $\{ (r_{i,j}, t_{i,j}) \mid 1 \leq j \leq m \}$, where $(r_{i,j}, t_{i,j})$ gives the $R$-time
and the $T$-time, respectively, for the job $J_i$ to be processed by the flowshop $M_j$. Of course,
constructing optimal schedules and developing good approximation algorithms on this more general model
become more challenging. We are currently working on this extended version of the scheduling model.

In many cases, a data request from a client is for a file, which consists of a number of data blocks
stored in either secondary or main memory. In a real system, it is possible that the disk-read and
network-transmission of the same file are executed in a pipeline manner in units of data blocks. Thus,
once a data block of a file is read entirely into the main memory, the block can be transmitted via
networks to the client even if some other data blocks for the file have not been read into the main
memory, yet. In particular, in such a model, preemption of processing data transmissions from servers
to clients becomes possible: after transmitting a few data blocks for a file $F$, a server may switch
to processing a different task, and come back later to continue transmitting the remaining data blocks
for the file $F$. Under this assumption, each data request from a client can be given by its size,
corresponding to the number of data blocks of the data, and the data request can be decomposed into a
continuous sequence of two-stage jobs, each corresponds to the disk-read and network-transmission of
a data block. Preemptions now are allowed on processing a data request from a client. However, frequent
preemptions for processing data requests should be avoided since restarting disk-read for a file will
require new disk search, which is significantly more time-consuming compared to reading a data block.
Therefore, when we study scheduling on this model, penalty on preemptions should be considered.

Scheduling with job precedences is very common in cloud computing practice. For example, a user who
wants to run a Microsoft application on a transparent computing platform may need from the cloud both
the code of the application as well as the code of Microsoft Windows software. However, the application
cannot be installed until the Windows software is installed on the client device \cite{transcomp}. As
a consequence, there is a need to study the scheduling problems under our model in which job
precedence is presented.

\end{document}